# PHONON RESIDUAL RESISTANCE OF PURE CRYSTALS

A.I. AGAFONOV

*National Research Center "Kurchatov Institute", Moscow, 123182 Russia*
*Agafonov_AIV@nrcki.ru*

Using the Boltzmann transport equation, we study the zero-temperature resistance of perfect metallic crystals of a finite thickness $d$ along which a weak constant electric field $E$ is applied. This resistance, hereinafter referred to as the phonon residual resistance, is caused by inelastic scattering of electrons heated by the electric field, with emission of long-wave acoustic phonons, and is proportional to $d^{-5}E^{-3}$. Consideration is carried out for Cu, Ag and Au perfect crystals with the thickness of about 1 cm, in the fields of the order of 1 mV/cm. Following the Matthiessen rule, the resistance of the pure crystals the thicknesses of which are much larger than the electron mean free path, is represented as the sum of both the impurity and phonon residual resistances. The condition on the thickness and field is found at which the low-temperature resistance of pure crystals does not depend on their purity and is determined by the phonon residual resistivity of the ideal crystals. The calculations are performed for Cu with a purity of at least 99.9999%.

## 1. Introduction

At low temperatures pure crystals possess the temperature-independent residual resistance[1,2,3]. Detailed studies[4] carried out for a number of good conductors such as Cu, Ag, Pt and Au, have shown that this resistance depends on the purity of the single crystals, the composition of impurities and crystal processing such as annealing[5]. If the crystal thicknesses are much larger than the impurity mean free path of electrons this resistance does not depend on crystal surface properties, and is determined only by the bulk properties[6,7]. For this reason it will be hereinafter referred to as the impurity residual resistance.

Any single crystal having ideal chemical and isotopic compositions, is inherent, however, non-ideality of the crystal lattice, which is caused by the zero-point vibrations of the atoms in the lattice sites. This quantum effect leads to a finite electron-phonon interaction even in the perfect crystal at zero temperature. From the experimental viewpoint the electric field in the crystal should be nonzero. Both of these factors lead to the fact that due to the work of the electric field the energy of electrons will always be sufficient to generate, at least, acoustic phonons. Then, even at zero temperature inelastic scattering of the field-heated electrons with emission of phonons should occur, and results in a finite value of the resistance that will below be referred to as the phonon residual resistance.

If the electron mean free path caused to this field-induced electron-phonon scattering, would be much less than the crystal thickness, the only defect of this thickness-limited perfect crystal, its surface, will not affect the conductivity. In this case the phonon residual resistance of the perfect crystal is determined only by its bulk properties.

The possibility of this electron-phonon contribution to residual resistivity was pointed out in the footnote[8], which was also noted that at zero temperature phonon scattering of electrons in the presence of electric field can lead to small nonlinear corrections to the electric field dependence of the current. As far as we know, the paper[9] is the only work in which the electron-phonon scattering contribution to metallic resistivity at 0 K has been studied. For finding the solution of the Boltzmann equation, the expansion of $f = f^0 + f^1$ was used with the assumption that $f^0 >> f^1$, where $f^0$ is the zero temperature Fermi distribution and $f^1$ is the perturbation distribution. As a result, a field-independent residual resistance was obtained. However, in the 0 K case (or at sufficiently low temperatures) this approximation is not valid because the field increase of the electron energies leads to the occupation of electronic states that were vacant initially. Hence, above the Fermi surface the field-dependent region which is significant for the conductivity, is appeared in which $f^1 >> f^0 = 0$.

Thus, even at zero temperature the perfect crystals of limited thicknesses can be characterized by a finite resistance. As far as we know, this phonon residual resistance which should be a nonlinear function of the electric field, has not been obtained previously. In pure crystals we can expect competition between the impurity and phonon residual resistances. However the phonon residual resistance has not yet been observed experimentally in pure crystals. We believe that this is due to a certain restriction on the field and the crystal thickness, at which the phonon residual resistance becomes greater than the impurity resistance in crystals of high purity.

The second section of this paper, using the Boltzmann transport equation, the phonon residual resistance of the perfect metallic crystals is derived in weak constant electric fields. The condition of smallness of the field is also discussed.

The third section, using the Matthiessen rule, the restriction on the field and the crystal thickness, at which the phonon residual resistance becomes greater than the impurity resistance in crystals of high purity, is found. Calculations are performed for copper crystals with a purity of at least 99.9999%. The Fröhlich electron-phonon coupling parameter can be determined from measurements of the phonon residual resistivity as a function of the field at a given thickness of the crystals.

## 2. Phonon residual resistance of thickness-limited perfect crystals

Let us assume that at temperature close to zero the metal crystal with the thickness $d$ is applied a constant electric field **E**. Then $\delta\varepsilon = eEd$ can be considered the maximum of the electron energy gain in the field $(e > 0)$. The same value corresponds to this energy gain of a hole below the Fermi surface. We suppose that $\delta\varepsilon$ is much smaller than the Debye temperature of the crystal,

$$\delta\varepsilon << k\Theta_D .  \qquad (1)$$

Then the scattering of electrons near the Fermi surface is accompanied by the emission of long-wavelength acoustic phonons with the energy

$$\hbar\omega_q = \hbar s q \leq \delta\varepsilon \ll k\theta_D, \qquad (2)$$

where $s$ is the speed of sound and $q$ is the phonon wave vector.

In the considered case the Boltzmann equation takes the form[8]:

$$e\mathbf{E}\frac{\partial f(\mathbf{k})}{\partial \mathbf{k}} = I[f(\mathbf{k})], \qquad (3)$$

where $f(\mathbf{k})$ is the electron distribution function in the field, $I[f(\mathbf{k})]$ is the collision integral that can be written as:

$$I[f(\mathbf{k})] = \int \frac{V_q^2 d\mathbf{q}}{(2\pi)^2} \{f(\mathbf{k})(1-f(\mathbf{k}-\mathbf{q}))\delta(\varepsilon_\mathbf{k} - \varepsilon_{\mathbf{k}-\mathbf{q}} - \hbar\omega_q) -$$
$$f(\mathbf{k}+\mathbf{q})(1-f(\mathbf{k}))\delta(\varepsilon_\mathbf{k} - \varepsilon_{\mathbf{k}+\mathbf{q}} + \hbar\omega_q)\}. \qquad (4)$$

Here $\varepsilon_\mathbf{k}$ is the electron energy with the electron wave vector $\mathbf{k}$, $V_q$ is the matrix element of the electron interaction with the long-wavelength acoustic phonons, the square of which is[10]:

$$V_q^2 = \lambda_0 \pi^2 \frac{\hbar^3 s q}{k_F m}, \qquad (5)$$

where $\lambda_0 < 1$ is the Fröhlich parameter, $m$ is the electron effective mass. In the second term in the curly brackets in (4) is convenient to use the replacement $\mathbf{q} \to -\mathbf{q}$.

Eq. (3) is valid only near the Fermi surface in a narrow layer $|\mu - \varepsilon_{\mathbf{k},\mathbf{k}-\mathbf{q}}| \leq \delta\varepsilon$. Then we can use $\varepsilon_\mathbf{k} = \mu + \hbar v_F(k - k_F)$, where $v_F$ и $k_F$ are the Fermi velocity and wave vector, respectively. Outside this layer as a boundary condition for the equation (3) we have the asymptotics:

$$f(\mathbf{k}) = \begin{cases} 1, \mu - \varepsilon_\mathbf{k} < \delta\varepsilon \\ 0, \varepsilon_\mathbf{k} - \mu > \delta\varepsilon \end{cases}. \qquad (6)$$

Solution of Eq. (3) - (4) is complicated by the fact that the collision integral contains the distribution function $f(\mathbf{k}-\mathbf{q})$, wherein the integration over the wave vectors of the phonons generated is performed.

As noted above, for finding the solution of the Boltzmann equation (3)-(4) the expansion of $f = f^0 + f^1(\mathbf{E})$ with the assumption $f^0 \gg f^1$ cannot be applied in the considered case. In fact, taking into account (6), one can see that $f^1 \gg f^0 = 0$ at $\mu \leq \varepsilon_k \leq \mu + \delta\varepsilon$. We use the following approach. In the narrow region $|\mu - \varepsilon_{\mathbf{k},\mathbf{k}-\mathbf{q}}| \leq \delta\varepsilon$ near the Fermi surface the electron wave vector is $k \approx k_F$, and the maximal wave vector of emitted phonons is

$$q_m = \delta\varepsilon / \hbar s = k_F \frac{\delta\varepsilon}{2\mu} \frac{v_F}{s}. \tag{7}$$

The typical ratio of the speed of sound to the Fermi velocity is $s/v_F \propto 10^{-3} \div 10^{-2}$. Therefore, for a sufficiently small ratio $\delta\varepsilon/\mu$ we can obtain the relation $q_m \ll k_F$ and substitute in (4) the replacement $f(\mathbf{k}-\mathbf{q}) \to f(\mathbf{k})$. Obviously, it can only be used under a certain restriction on the field.

To find this restriction, in the collision integral (4) we substitute the expansion $f(\mathbf{k}-\mathbf{q}) \cong f(\mathbf{k}) - \mathbf{q}\frac{\partial f(\mathbf{k})}{\partial \mathbf{k}}$. As a result, we have:

$$I[f(\mathbf{k})] = \frac{1}{(2\pi)^2 \hbar v_F} f(\mathbf{k})(1-f(\mathbf{k})) \int V_q^2 d\mathbf{q} \left[ \delta(k-|\mathbf{k}-\mathbf{q}|-\frac{s}{v_F}q) - \delta(k-|\mathbf{k}-\mathbf{q}|+\frac{s}{v_F}q) \right] +$$

$$\frac{1}{(2\pi)^2 \hbar v_F} f(\mathbf{k}) \frac{\partial f(\mathbf{k})}{\partial \mathbf{k}} \int V_q^2 \mathbf{q} d\mathbf{q} \left[ \delta(k-|\mathbf{k}-\mathbf{q}|-\frac{s}{v_F}q) - \delta(k-|\mathbf{k}-\mathbf{q}|+\frac{s}{v_F}q) \right] + \tag{8}$$

$$\frac{1}{(2\pi)^2 \hbar v_F} \frac{\partial f(\mathbf{k})}{\partial \mathbf{k}} \int V_q^2 \mathbf{q} d\mathbf{q} \, \delta(k-|\mathbf{k}-\mathbf{q}|+\frac{s}{v_F}q).$$

Introducing the angle $\theta$ between the vectors $\mathbf{k}$ and $\mathbf{q}$, from (8) we conclude that the wave vector of the emitted phonons is almost perpendicular to the electron wave vector, as $\cos\theta = \frac{s}{v_F} \mp \frac{q}{2k}(1-\frac{s^2}{v_F^2}) \ll 1$. Taking into account of (7), the right-hand side of (8) is rewritten as:

$$I[f(\mathbf{k})] = -\frac{1}{\pi \hbar v_F k} \frac{s}{v_F} f(\mathbf{k})(1-f(\mathbf{k})) \int_0^{q_m} V_q^2 q^2 dq +$$

$$\frac{1}{\pi \hbar v_F} \frac{s}{v_F} f(\mathbf{k}) \frac{\partial f(\mathbf{k})}{\partial \mathbf{k}} \frac{\mathbf{k}}{k} \int_0^{q_m} V_q^2 q^2 dq \left( 1 - \frac{q^2}{2k^2}\left(1-\frac{s^2}{v_F^2}\right) \right)$$

$$-\frac{1}{2\pi \hbar v_F} \frac{\partial f(\mathbf{k})}{\partial \mathbf{k}} \frac{\mathbf{k}}{k} \int_0^{q_m} V_q^2 q^2 dq \left(1+\frac{s}{v_F}\frac{q}{k}\right)\left(\frac{s}{v_F} - \frac{q}{2k}\left(1-\frac{s^2}{v_F^2}\right)\right). \tag{9}$$

In (9) the terms, containing $sq/kv_F$, $q^2/k^2$ и $s^2/v_F^2$ under the integrals, can be omitted. Then, the last two terms on the right hand side of (9) are reduced to the form:

$$\frac{1}{\pi \hbar v_F} \frac{s}{v_F} \left( f(\mathbf{k}) - \frac{1}{2} \right) \frac{\partial f(\mathbf{k})}{\partial \mathbf{k}} \frac{\mathbf{k}}{k} \int_0^{q_m} V_q^2 q^2 dq + \frac{1}{4\pi \hbar v_F} \frac{\partial f(\mathbf{k})}{\partial \mathbf{k}} \frac{\mathbf{k}}{k^2} \int_0^{q_m} V_q^2 q^3 dq.$$

These two terms must be small compared to the first term on the right side (9). As a result, taking into account that $f(\mathbf{k}) \leq 1$, we obtain the condition:

$$f(\mathbf{k})(1-f(\mathbf{k})) \gg \frac{1}{2}\left(1+\frac{1}{5}\frac{\delta\varepsilon}{\mu}\left(\frac{v_F}{s}\right)^2\right)\left|\mathbf{k}\frac{\partial f(\mathbf{k})}{\partial \mathbf{k}}\right|. \quad (10)$$

Here $\mathbf{k}$ is any wave vector lying within the region $|\mu - \varepsilon_\mathbf{k}| \leq \delta\varepsilon$.

As we show below, the condition (10) is reduced to the restriction on the electric field applied to the crystal sample. In the case (10), the kinetic equation (3) - (4) with account of (5) takes the form:

$$e\mathbf{E}\frac{\partial f(\mathbf{k})}{\partial \mathbf{k}} = -\frac{\pi\lambda_0}{2^5}\frac{\delta\varepsilon^4}{\mu^3}\left(\frac{v_F}{s}\right)^2\frac{k_F}{k}f(\mathbf{k})(1-f(\mathbf{k})). \quad (11)$$

The ratio $k_F/k$ can be replaced by 1 with high precision.

Crystals of Cu, Ag, and Au have a face-centered cubic lattice. Let the field $\mathbf{E}$ coincides with the direction [111] along the diagonal of the cube. Then (11) can be rewritten as:

$$\left(\frac{\partial}{\partial k_x} + \frac{\partial}{\partial k_y} + \frac{\partial}{\partial k_z}\right)\ln\frac{f(\mathbf{k})}{1-f(\mathbf{k})} = -\xi, \quad (12)$$

where

$$\xi = \frac{\sqrt{3}\pi\lambda_0}{2^5}\left(\frac{\delta\varepsilon}{\mu}\right)^3\left(\frac{v_F}{s}\right)^2 d. \quad (13)$$

Taking into account (6), from Eq. (12) we find the electron distribution function:

$$f(\mathbf{k}) = \begin{cases} 1, & k \leq k_F - \dfrac{\delta\varepsilon}{\hbar v_F} \\ \dfrac{1}{1+\exp(\xi(\mathbf{ek}-k*))}, & |k_F - k| \leq \dfrac{\delta\varepsilon}{\hbar v_F} \\ 0, & k \geq k_F + \dfrac{\delta\varepsilon}{\hbar v_F} \end{cases} \quad (14)$$

Here the unit vector $\mathbf{e} = \mathbf{E}/E$, $k*$ is the parameter determined by the condition of electrical neutrality of the crystal:

$$\sum_\mathbf{k}[f(\mathbf{k}) - \Theta(\mu - \varepsilon_\mathbf{k})] = 0, \quad (15)$$

where $\Theta$ is the Heaviside step function.

Now, for the above-mentioned good conductors Cu, Ag and Au we perform calculations the required fields for which one can use the distribution function (14). Taking into account (13) and (14), from (10) we obtain the restriction on the electric field in the metal crystal:

$$\frac{2^6}{\sqrt{3}\pi\lambda_0}\left(\frac{s}{v_F}\right)^2\frac{1}{k_F d} \gg \left(\frac{\delta\varepsilon}{\mu}\right)^3 + \frac{1}{5}\left(\frac{v_F}{s}\right)^2\left(\frac{\delta\varepsilon}{\mu}\right)^4. \quad (16)$$

Let the crystal thickness $d = 1$ cm and the Fröhlich parameter $\lambda_0 = 1/3$ (as known[8], this dimensionless parameter $\lambda_0 < 1$ for the metals).

For Cu the Fermi energy $\mu \cong 7.0$ eV, the electron effective mass $m \cong m_e$, velocity of longitudinal sound waves $s \cong 4.6*10^5$ cm/s. Consequently, we obtain that the Fermi velocity $v_F \cong 1.57*10^8$ cm/s and wave vector $k_F \cong 1.35*10^8$ cm$^{-1}$. Then, the condition (16) leads to the restriction on the voltage drop across the crystal $\delta\varepsilon < 0.63$ meV. Hence, the electric field should be $E < 0.63$ mV/cm in the copper sample thickness $d = 1$ cm.

For silver $\mu \cong 5.5$ eV, $m \cong m_e$ and $s \cong 3.6*10^5$ cm/s. Consequently, the Fermi velocity and wave vector are equal to $v_F \cong 1.4*10^8$ cm/s и $k_F \cong 1.2*10^8$ cm$^{-1}$, respectively. For the same the Fröhlich parameter and crystal thickness the restriction is $E < 0.45$ mV/cm.

For Au $\mu \cong 5.5$ eV, $m \cong 1.1 m_e$, $s \cong 3.2*10^5$ cm/s and, accordingly, the inequality (16) gives $E < 0.42$ mV/cm.

Thus, the electron distribution function is given by (14) for fields satisfying the above constraints. Comparing (14) with the equilibrium Fermi distribution, we conclude that the phonon scattering of the field-heated electrons changes the **k**-distribution of electrons near the Fermi surface in the layer $2\delta\varepsilon = 2eEd$. To avoid smearing of this feature, the latter imposes a limitation on the crystal temperature $kT << \delta\varepsilon$.

Using (14) one can find the parameter $k_*$ and the current density in the crystal. The equation for $k_*$ is

$$\int_{k_F - \frac{\delta\varepsilon}{\hbar v_F} \leq k \leq k_F} d\mathbf{k} = \int_{|k - k_F| \leq \frac{\delta\varepsilon}{\hbar v_F}} \frac{d\mathbf{k}}{1 + \exp(\xi(\mathbf{ek} - k_*))}, \quad (17)$$

and the current is defined as:

$$J = -\frac{e\hbar}{4\pi^3 m} \int_{|k_F - k| \leq \frac{\delta\varepsilon}{\hbar v_F}} \frac{(\mathbf{ek}) d\mathbf{k}}{1 + \exp(\xi((\mathbf{ek}) - k_*))}. \quad (18)$$

The integrals (17) and (18) over the spherical shell are reduced to one-dimensional integrals[11]. As a result, Eq. (17) takes the form:

$$4k_F^2 \frac{\delta\varepsilon}{\hbar v_F}\left(1 - \frac{\delta\varepsilon}{2\mu}\right) = \int_{-k_F - \frac{\delta\varepsilon}{\hbar v_F}}^{k_F + \frac{\delta\varepsilon}{\hbar v_F}} \frac{(k_F + \frac{\delta\varepsilon}{\hbar v_F})^2 - t^2}{1 + \exp(\xi(t - k_*))} dt - \int_{-k_F + \frac{\delta\varepsilon}{\hbar v_F}}^{k_F - \frac{\delta\varepsilon}{\hbar v_F}} \frac{(k_F - \frac{\delta\varepsilon}{\hbar v_F})^2 - t^2}{1 + \exp(\xi(t - k_*))} dt, \quad (19)$$

where the term $\delta\varepsilon^2 / 12\mu^2$ in the round bracket on the left side of (19) is omitted, and the current (18) is rewritten as:

$$J = -\frac{e\hbar}{4\pi^2 m} \left[ \int_{-k_F - \frac{\delta\varepsilon}{\hbar v_F}}^{k_F + \frac{\delta\varepsilon}{\hbar v_F}} \frac{(k_F + \frac{\delta\varepsilon}{\hbar v_F})^2 - t^2}{1 + \exp(\xi(t - k_*))} t dt - \int_{-k_F + \frac{\delta\varepsilon}{\hbar v_F}}^{k_F - \frac{\delta\varepsilon}{\hbar v_F}} \frac{(k_F - \frac{\delta\varepsilon}{\hbar v_F})^2 - t^2}{1 + \exp(\xi(t - k_*))} t dt \right]. \quad (20)$$

According to (13), the value $\xi \propto E^3$. For the restriction (16) we have $\xi k_F << 1$ and $\xi \delta\varepsilon / \hbar v_F << 1$. Then, in Eqs. (19) and (20) we can use the expansion $(1+\exp(\xi(t-k*)))^{-1} = \frac{1}{2} - \frac{1}{4}\xi(t-k*)$. As a result, from Eq. (19) we obtain: $\xi k* = -\frac{\delta\varepsilon}{\mu} \ll 1$.

Finally, the current is given by:

$$J = \frac{\lambda_0}{8\sqrt{3}\pi} \frac{e^5 E^4 d^5}{\hbar^4 v_F s^2}. \qquad (21)$$

Note that for the perfect crystal the current is proportional to $J \propto d^5 E^4$. If the crystal thickness $d \to \infty$, then $J \to \infty$ at a finite value of the field $E$. In the limit $E \to 0$ the current vanishes. In this case the field heating of electrons does not occur and, respectively, this electron-phonon scattering becomes impossible.

From (21) we obtain the expression for the phonon residual resistance of the perfect crystal ($\rho_{pcr} = E/J$):

$$\rho_{pcr} = \frac{4\sqrt{3}}{\lambda_0} R_K \frac{\hbar^3 v_F s^2}{e^3 U^3 d^2}, \qquad (22)$$

where $R_K$ is the von Klitzing constant ($R_K = 25.8 K\Omega$), $U = Ed$ is the voltage drop across the crystal. Note that the phonon residual resistance is proportional to $d^{-5} E^{-3}$. Calculations of the phonon residual resistance will be given in the next section.

## 3. Residual resistance of pure crystals

Pure crystals can be characterized by both the impurity and phonon residual resistance. For a comparative analysis of these two contributions we use the Matthiessen rule, according to which the resistivity is the sum of the partial resistances for each scattering channel: $\rho = \rho_0 + \rho_{pcr}$, where $\rho_{pcr}$ is determined by (22) and $\rho_0$ is the impurity residual resistance which is proportional to the concentration of impurities.

Now, we find the condition at which the phonon residual resistance of the perfect crystals can be experimentally observed for pure crystals. According to (22), it is possible if

$$E^3 d^5 < \frac{4\sqrt{3}}{\lambda_0} R_K \frac{\hbar^3 v_F s^2}{e^3 \rho_0}, \qquad (23)$$

where the crystal thickness $d$ should be much larger than the impurity free mean path of electrons: $d \gg \hbar k_F /(e^2 \rho_0 n)$. Here $n$ is the electron density of the crystal.

Note that under the inequality (23) the resistance can be independent on the crystal purity, and is determined by (22) as though the crystal is perfect.

The review article[4] is collected the well-known experimental data on the electrical resistivity of pure crystals of Cu, Ag, Pd, Au. For us the temperature-independent impurity residual resistances are important. For copper with the purity of 99.999% the recommended value of the residual resistance $\rho_0 \cong 2*10^{-9} \Omega cm$, as shown by the thick curve in Fig.5 in the paper[4]. Copper with a purity of 99.9999% has the resistance of about one order of magnitude lower than that presented above (see the curve 279 in Fig. 5). The value $\rho_0 \cong 1.2*10^{-10} \Omega cm$ can be reached in the more pure copper (the curve 185 in Fig. 5).

Given (22), the total resistance can be presented in the form:

$$\rho = \rho_0 \left(1 + CE^{-3} d^{-5}\right). \quad (24)$$

In (24) for the value $\rho_0 \cong 1.2*10^{-10} \Omega cm$ we obtained $C = 0.042$, where $d$ and $E$ should be substituted in units of cm and of mV/cm, respectively. Assuming the electron density $n = 8.45*10^{22}$ cm$^{-3}$ for Cu, the crystal thickness should be $d >> \hbar k_F /(e^2 \rho_0 n) = 0.06$ cm. For each value of $d$ the electric field is limited by the condition (16). The required temperatures are estimated based on the ratio $T = \delta\varepsilon/3k$.

Fig. 1 presents our calculated data for copper with the impurity residual resistivity $\rho_0 = 1.2*10^{-10} \Omega cm$. The curve 1 in Fig. 1 corresponds to equality of the right and left sides of (16). In the region below this curve the inequality (16) is satisfied and, accordingly, the phonon residual resistance is given by (22).

The curve 2 in Fig. 1 demonstrates the crystal thickness dependence of the electric field at which the phonon residual resistance is equal to the impurity resistance. According to (24), the dependence has the form $E = 0.042/d^{5/3}$, and the crystal resistance $\rho = 2\rho_0$. Note that for not too thin samples the curve 2 lies below the curve 1, as shown in Fig. 1. Starting from these points on the curve 2, the decrease of the electric field leads to the sharp increase in the crystal resistivity.

On the inset of Fig.1, for the 0.5 cm thick crystal the two curves marked by 3, represent the region of electric field (solid line) and the required temperatures (dashed line) in which the phonon residual resistance changes from $\rho_0$ to $10\rho_0$. In this region, the sharp increase in the resistance of the crystal is predicted with decreasing of the electric field. Here, taking into account of (22), the Fröhlich electron-phonon coupling parameter ($\lambda_0$) can be found for the metal.

For the 1 cm thick crystal this region corresponding to the tenfold increase in the resistance, is characterized by significantly lower values of the fields and temperatures, as demonstrated by the two curves marked by 4 in the inset of Fig.1.

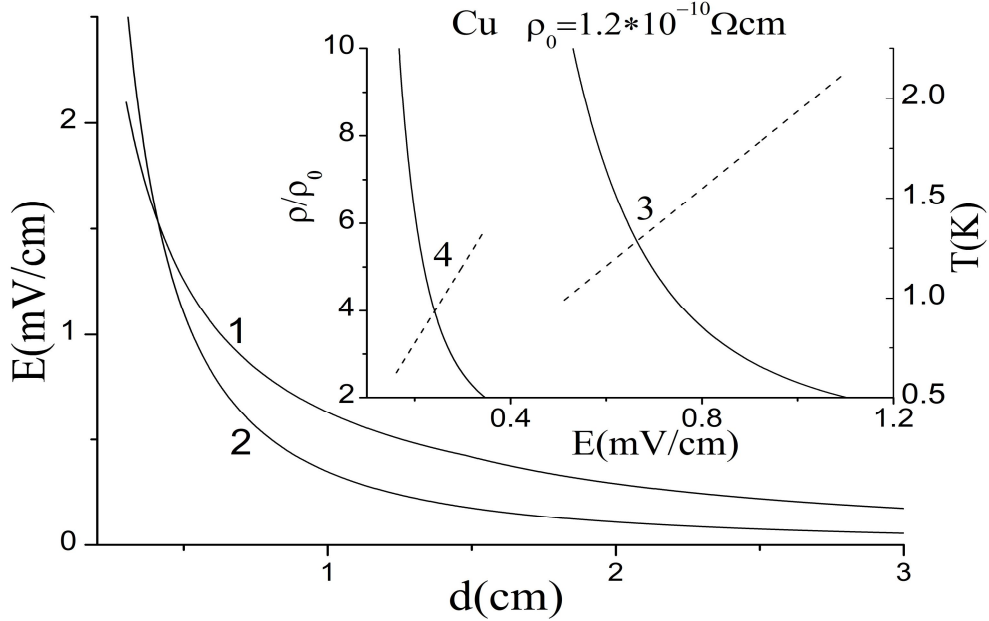

Fig. 1. Explanations in the text.

For Cu with the impurity residual resistance $\rho_0 \cong 2*10^{-9}\,\Omega cm$ the value of $C = 0.0025$, and the crystal thickness should be $d \gg 3.6*10^{-3}$ cm. We conclude that for the samples with the same thickness more favorable conditions for experimental observation of the phonon residual resistance are realized for the more pure samples of copper. Using the data[4], similar calculations are easily carried out for the other materials mentioned above.

### 4. Conclusion

It has been shown that even at zero temperature the perfect crystals of limited thickness possess the finite resistance, which is due to phonon inelastic scattering of electrons heated by the electric field. This resistance is determined by the bulk characteristics of the ideal crystal, in which the electron-phonon interaction is caused by zero-point vibrations of the atoms in the perfect crystal lattice sites. The main result is the expression for the phonon residual resistance (22) which is valid for the electric fields satisfying the inequality (16).

Improvement of the electrical conductivity of pure crystals is associated, as a rule, with the increasing of their purity. It results in the decreasing of the impurity residual resistivity only, but does not affect the phonon residual resistance. Using the Matthiessen rule, we have found the condition (23) at which the resistance of pure crystals does not depend on their purity, and is determined by the phonon residual resistance, as though the pure crystal would be perfect.

*A.I. Agafonov*

**Acknowledgments**

I am thankful to Prof. S.V. Sazonov for helpful discussions.

**References**

1. I.M. Lifshitz, M.I. Kaganov, *Sov. Phys. Usp.* **8**, 805 (1966).
2. Yu. Kagan, A.P. Zhernov, *JETP* **26,** 999 (1968).
3. P.L. Rossiter. *The electrical resistivity of metals and alloys* (Cambridge Solid State series, Cambridge University Press, 1987).
4. K.A.Matula, *J. Chem. Ref. Data* **8**, 1147 (1979).
5. A. Kurosaka, H. Tominaga, H. Osanai, *Advances in Cryogenic Engineering Materials* **36**, 749 (1990).
6. V. Kuckhermann, G. Thummes, H.H. Mende, *J. Phys. F: Met. Phys.* **15**, L153 (1985).
7. Yu.P. Gaidukov, *Sov. Phys. Usp.* **27,** 256 (1984).
8. A.A. Abrikosov. *Fundamentals of the theory of metals*. (North Holland, Amsterdam, (1988).
9. R.L. Liboff, G.K. Schenter, *Phys. Rev.* B **54**, 16591 (1996).
10. G.M. Eliashberg, *Sov. Phys. JETP* **11**, 696 (1960)].
11. A.P. Prudnikov, Yu. A. Brychkov, and O.I. Marichev. *Integrals and series.* (Fizmatlit, Moscow, 2002) vol. 1, C. 479 (in Russian).